\documentclass[a4paper,11pt]{article}
\usepackage{jinstpub} 

\usepackage{xcolor}
\usepackage[utf8]{inputenc}
\usepackage[T1]{fontenc}
\usepackage{textcomp}
\usepackage{siunitx}
\usepackage{amsmath}
\usepackage{amsthm}
\usepackage{upgreek}
\usepackage{eurosym}
\usepackage{xspace}
\usepackage{textpos}
\usepackage{textcomp}
\usepackage{multirow}
\usepackage{array}
\usepackage{subcaption}

\usepackage{float}

\usepackage{lineno}

\def\um{\si{\micro\m}\xspace}

\def\Sr{$^{90}$Sr\xspace}

\def\B{$^{10}\mathrm{B}$\xspace}

\def\B11{$^{11}\mathrm{B}$\xspace}
\def\Bi{$^{207}\mathrm{Bi}$\xspace}

\def\Sr{\ensuremath{^{90}}Sr\xspace}
\def\Y{\ensuremath{^{90}}Y\xspace}
\def\Zr{\ensuremath{^{90}}Zr\xspace}

\def\allpix{Allpix${^2}$\xspace}

\graphicspath{{img/}}

\title{Study of electron tracks in Timepix3 detector at kinetic energies of 1 and 1.5~MeV}

 \author[a]{Babar Ali}
 \author[a,b]{Zden\v ek Kohout}
 \author[a,1]{Hugo Natal da Luz,\note{Corresponding author.}}
 \author[a]{Rudolf S\'ykora}
 \author[a]{Tom\'a\v s S\'ykora}
 \affiliation[a]{Institute of Experimental and Applied Physics, Czech Technical University in Prague,\\Husova 5/240, 110 00 Prague 1, Czech Republic}
 \affiliation[b]{Faculty of Mechanical Engineering, Czech Technical University in Prague,\\Technická 4, 160 00 Prague 6, Czech Republic}
 
\emailAdd{hugo.nluz@cvut.cz}

\abstract{We report on measurements of 1 and 1.5\,MeV
monoenergetic electrons with a Timepix3-based detector using
a 0.5\,mm thick silicon sensor.  A \Sr $\beta$-emitting
radioisotope was used as the source of electrons, and a
monochromator equipped with an adjustable magnetic field was
employed to only pass electrons of desired energy into the
detector. We provide experimental results of
deposited-energy spectrum in the sensor and linearity of
detected tracks.  Alongside with the experiment, the whole
system has been modelled in software and a Monte Carlo
Geant4 / \allpix simulation of the experiment has been carried out.
Generally, we find a good agreement between the two.}


\arxivnumber{2411.19081} 

\begin{document}
\maketitle
\flushbottom

\section{Introduction}
\label{sec:intro}

Motivated by a wish to remeasure the so-called ATOMKI
anomaly reported in 2016 \cite{Attila1}, we are building a
detector suitable for measuring energies and directions of
electrons and positrons, of typical energy about 9\,MeV,
emerging from a nuclear reaction taking place in a target
hit by accelerated protons. The innermost part of the
detector will consist of a ring of six Timepix3-based (TPX3)
\cite{Poi14} silicon pixel detectors. Since there is no
suitable source of electrons (positrons) of the given
energies available to us, we, to an extent, rely on
simulations to predict the detector behaviour. Nonetheless,
even the simulation framework itself needs to be tested,
preferrably against experimentally verifiable data. We
therefore conducted several measurements with simplified
setups utilizing accessible electron sources, though, of
lower energies. On one hand we believe that many conclusions
drawn from such experiments will apply to the final
(ATOMKI-like) experiment, too. On the other, the results may
potentially be interesting on their own right. Thus, in this
work we describe measurements with a single TPX3 detector of
monoenergetic semi-relativistic electrons with kinetic
energy of 1 and 1.5\,MeV, obtained from a \Sr radioactive
source by means of selection by an adjustable magnetic
monochromator. The experiments are complemented by thorough
simulations and the results are mutually compared.

The TPX3 is a read-out chip for both semiconductor and
gas-filled detectors. Developed within the Medipix
collaboration~\cite{Poi14}, the chip is a square matrix of
$256\times 256$ pixels, each of $55\,\um\times55\,\um$ size,
hence with a total active area of
$14\times14$\,\si{\square\mm}. The chip is typically
bump-bonded to a suitable sensor and collects the charge
left behind ionizing radiation. In the so-called data-driven
mode, which we employed, the information about an activated
pixel is made available without delay and independently of
other pixels.  Each pixel measures both the time-of-arrival
(ToA) and the time-over-threshold (ToT) of the detected
signal, with a precision of 1.6\,ns. ToT reflects the energy
deposited in the pixel.

This work used a TPX3 detector equipped with a 0.5-mm-thick
silicon sensor. The detector was read by the so-called
Katherine readout \cite{Bur17} and the acquired data were
transferred through ethernet to a personal computer for
further analysis.

It has been shown \cite{Goh16,Goh19} that electrons (and
positrons) of the mentioned energies (i.e., of the order of
1\,MeV) are frequently multiply scattered within the
detector sensor, leading to wiggly tracks which hinders
determination of the particle incoming direction.  On the
other hand, it is these processes that allow full-energy
deposition in a sensor with thickness smaller than would
correspond to electron range calculated solely from the
sensor-material linear stopping power. It is the aim of this
paper to provide quantitative evidence of such effects,
i.e., show experimental results of measured electron energy
and linearity of tracks.  At the same time we want to
document that our Monte Carlo simulations of the experiment
lead to its faithful representation, the importance of which
lies in certain justification of using the developed
simulation framework also later for our further experiments.
Last, while setting up the simulation we built a detailed
model of the monochromator, which may turn useful for other
researchers using the device.

\section{Experimental setup --- Electrons from \Sr source}
\label{sec:exKralupy}

\begin{figure}
    \centering
    \includegraphics[width=.7\textwidth]{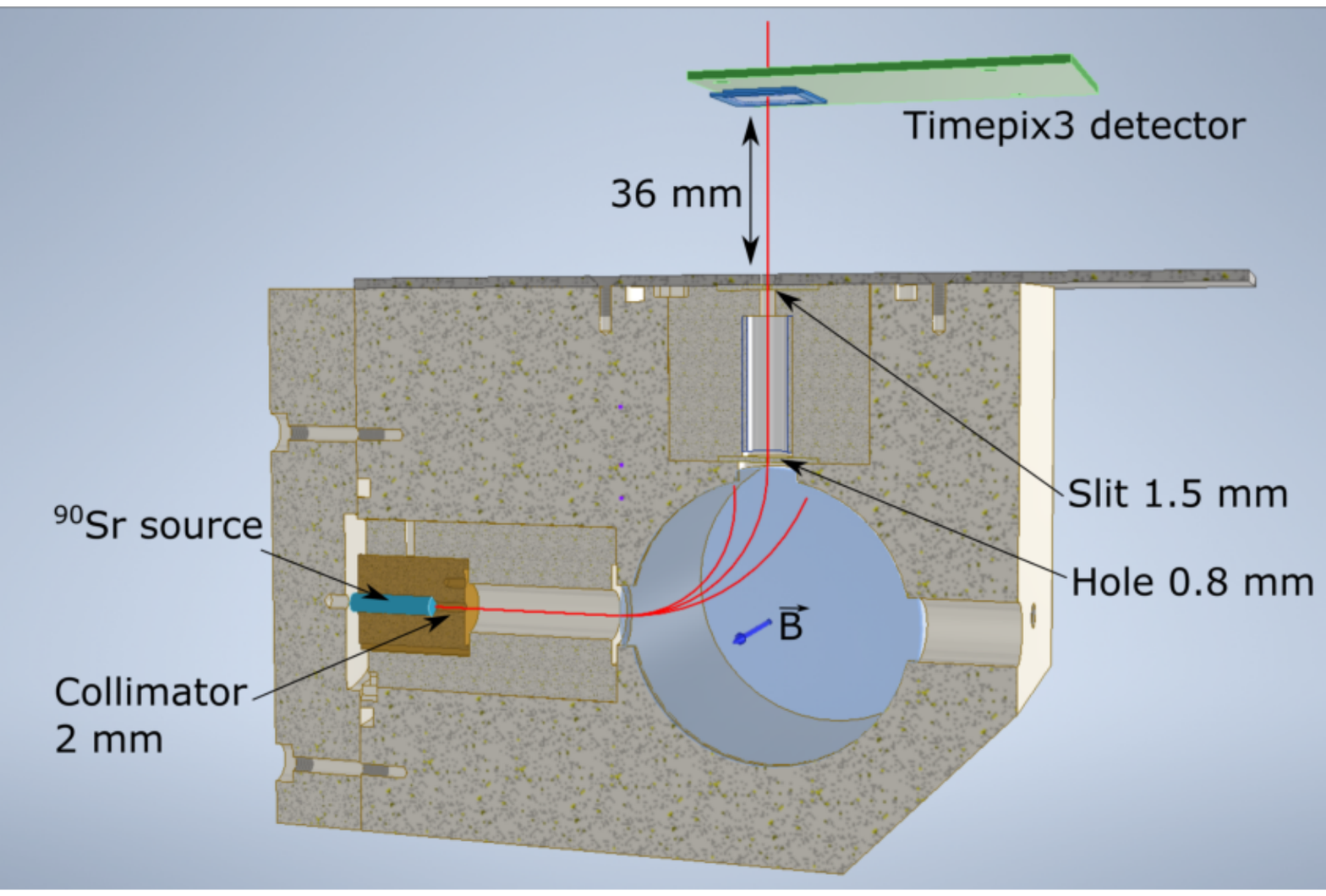}
    \caption{Schematics of the experimental setup;
      the electromagnet parts have been left out for clarity.}
    \label{fig:schKral}
\end{figure}

A monoenergetic electron beam is obtained from a $^{90}$Sr
$\beta$ source%
\footnote{
Unstable \Sr decays to \Y via $\beta^-$ decay, with a Q$_\beta$
of 0.55 MeV.  Subsequently, \Y decays to stable \Zr via
$\beta^-$ decay, with a Q$_\beta$ value of 2.28 MeV.
}
supplemented with a tuneable magnetic monochromator.  Figure
\ref{fig:schKral} shows a CAD drawing of the setup.
Electrons emitted from \Sr enter a region with magnetic field
provided by an iron-core electromagnet. The current through
its coils is adjustable and only electrons of certain energy
make it through the depicted apertures to the outside world.
The chamber of the monochromator is kept at 10-mbar
pressure, and the electrons exit the monochromator through a
12-\um-thick mylar window. The setup produces electrons in
useful amounts from 0.4 and 1.8\,MeV with the intensity
peaking at 1\,MeV and dropping to half at 0.5 and 1.5\,MeV.
Beam energy resolution varies from approximately 10\% at 150
keV to 3\% at 1.5 MeV. Reference~\cite{Sin17} provides
further technical details.

\begin{figure}
    \centering
    \begin{subfigure}{0.46\textwidth}       
    \includegraphics[width=\textwidth]{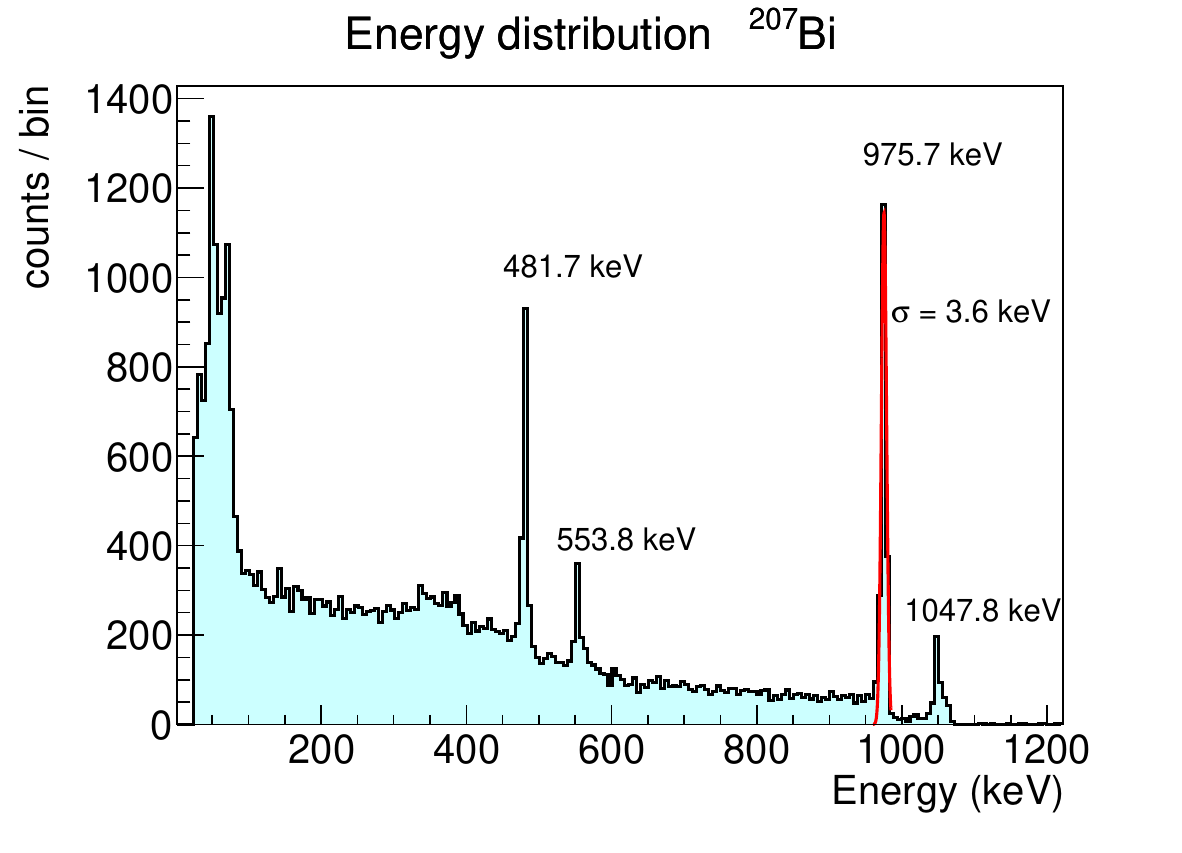}
    \caption{}
    \label{fig:Bi_a}
   \end{subfigure}
    \begin{subfigure}{0.46\textwidth}       
    \includegraphics[width=\textwidth]{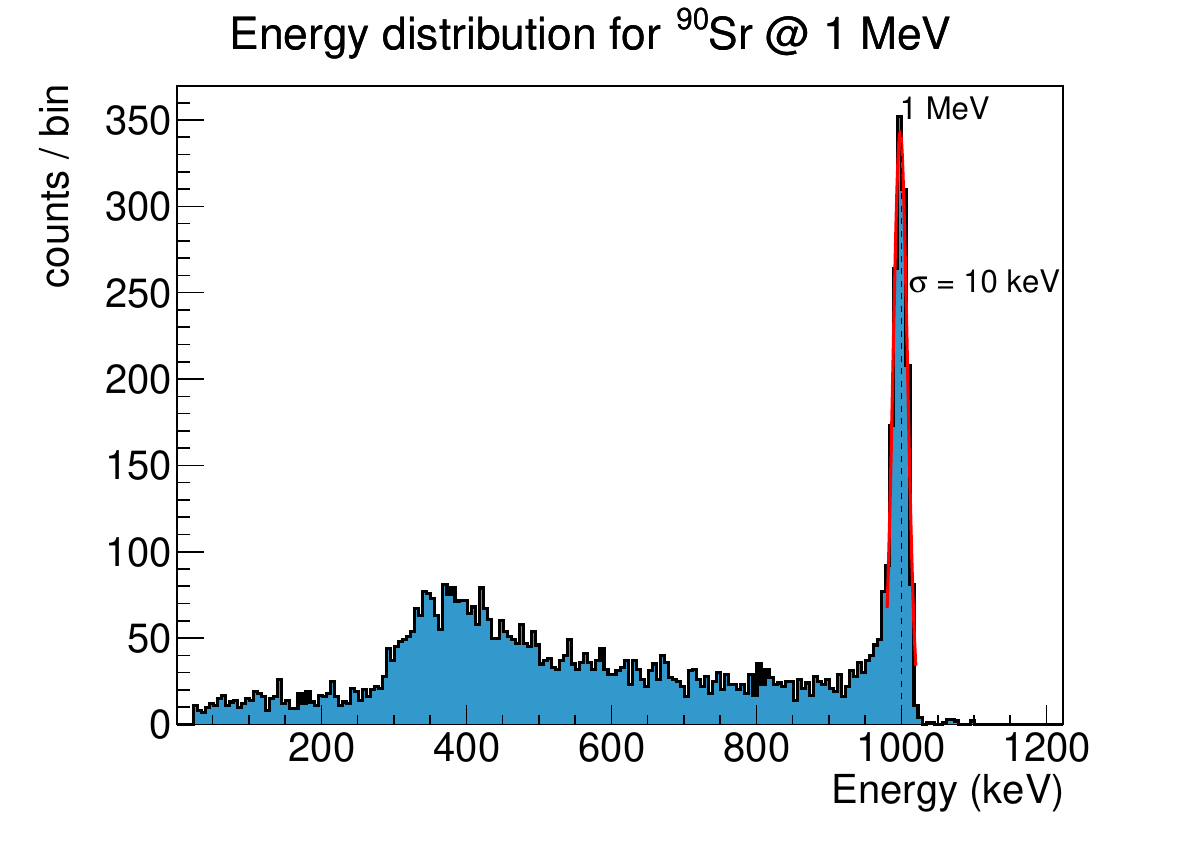}
    \caption{}
    \label{fig:Bi_b}
   \end{subfigure}
    \caption{Electron-energy distribution measured by a 1-mm-thick
      silicon diode for electrons originating from two different sources: 
      (a) \Bi,
      (b) \Sr source with monochromator set to emit 1-MeV particles.}
    \label{fig:Bi}
\end{figure}

The calibration of the electron source (i.e., the relation
between outgoing particle energy and electromagnet current)
was checked with a 1-mm-thick silicon-diode detector that
had been pre-calibrated using the 481.7- and 976-keV
conversion electrons emitted from a \Bi source \cite{Kon11}.
Figure \ref{fig:Bi_a} shows the \Bi electron-energy spectrum
measured with the diode, Fig.~\ref{fig:Bi_b} then depicts
the electron-energy spectrum, measured with the same
detector, from the calibrated \Sr source set to output 1-MeV
particles.  (1-MeV electrons generally pass through the 1-mm
silicon diode. If viewed as `heavy' particles, they would
most probably deposit about 350\,keV in the sensor, which is
reflected in the broad Landau-like peak in
Fig.~\ref{fig:Bi_b}; since they are not heavy and therefore
scatter frequently, some electrons still loose all their
energy in the sensor.)

Since the real energy of 976-keV electrons is well-defined,
the respective peak width in Fig.~\ref{fig:Bi_a} corresponds
to the detector resolution; the Gaussian fit yields
the sigma of about 4 keV. From this and from the
measured 10-keV Gaussian sigma-width of the 1-MeV peak in
Fig.~\ref{fig:Bi_b} one can estimate the true energy spread
of the electrons leaving the \Sr-based source to be of about
9\,keV sigma, i.e., about 20\,keV FWHM (2\% at 1\,MeV), in
accordance with \cite{Sin17}.

In experiments described in this work a TPX3 detector was
placed at a fixed distance of 36 mm from the monochromator
exit window, and electrons with energies of 1\,MeV and 1.5\,MeV
were used.

\section{Simulations}
\label{sec:simul}

A 3D model of the complete experimental setup,
including the parts of the monochromator and the TPX3
detector, has been built using CAD software and described in
the geometry description markup language (GDML).
The GDML description, besides the geometry, also specifies
the material of each part of the setup.

Geant4~\cite{Geant4}, a Monte Carlo-based toolkit, takes the
GDML model and simulates propagation of electrons emitted
from the \Sr source through the monochromator and into and
within the TPX3 sensor. It takes care of any secondary
particles involved, namely, it reports about creation of
electron--hole pairs in the TPX3 sensor brought about by the
passage of the primary electron.%
\footnote{
Geant4's \texttt{QGSP\_BERT\_EMZ} physics list has been used.
}
For simplicity, the \Sr source is modelled to emit electrons
isotropically, the cavities of the monochromator are
represented by vacuum, and the magnetic field in the
monochromator chamber is taken to be homogeneous.  The
strength of the magnetic field is adjusted to have electrons
of wanted energy hit the detector.

The information about created electron--hole pairs in the
sensor, as provided by Geant4, is taken over by the \allpix
\cite{Allpix2a,Allpix2b} program, which simulates the
evolution of the charges within the sensor, their
collection, and the generation and digitization of the
signal eventually provided by the TPX3 detector. Thus,
\allpix output parallels that of a real experiment, reporting
when (ToA) and what pixels were activated, as well as the
energy deposited in the individual pixels (ToT).  The
\allpix program is intimately aware of the TPX3 detector
internals. Furthermore it takes a set of parameters
specifying current operating conditions, such as those
listed in Table \ref{tab:allpix_param}; the values there
correspond to the real conditions of our detector during the
data taking.

The geometry of the electron monochromator used in our
simulations is based on existing documentation.  However, it
is known that the monochromator construction did not adhere
exactly to the documented specifications.  Since the device
is sealed, it was not possible to verify the exact
dimensions and shapes of the slits and collimators.  The
final model is therefore also partially based on results
from a series of attempts to reproduce the experimental
data.

\begin{table}[htbp]
  \caption{Operating parameters used in the \allpix simulation for the TPX3 detector.}
  \label{tab:allpix_param}
  \begin{center}
      \begin{tabular}{l|c}
      \hline \hline
        Bias voltage & 210 V \\
        Depletion voltage & 100 V \\
        Temperature & 352 K \\
        Electronic noise & 100 e$^{-}$ \\
        Threshold & (1248 $\pm$ 35)~e$^{-}$ \\
        Integration time & 100 ns \\
        \hline \hline
      \end{tabular}
  \end{center}
\end{table}

\section{Results}
\label{sec:results}

\subsection{Energy distribution}
\label{ssec:edist}

\begin{figure}[htbp]
  \centering 
    \includegraphics[width=0.5\textwidth]{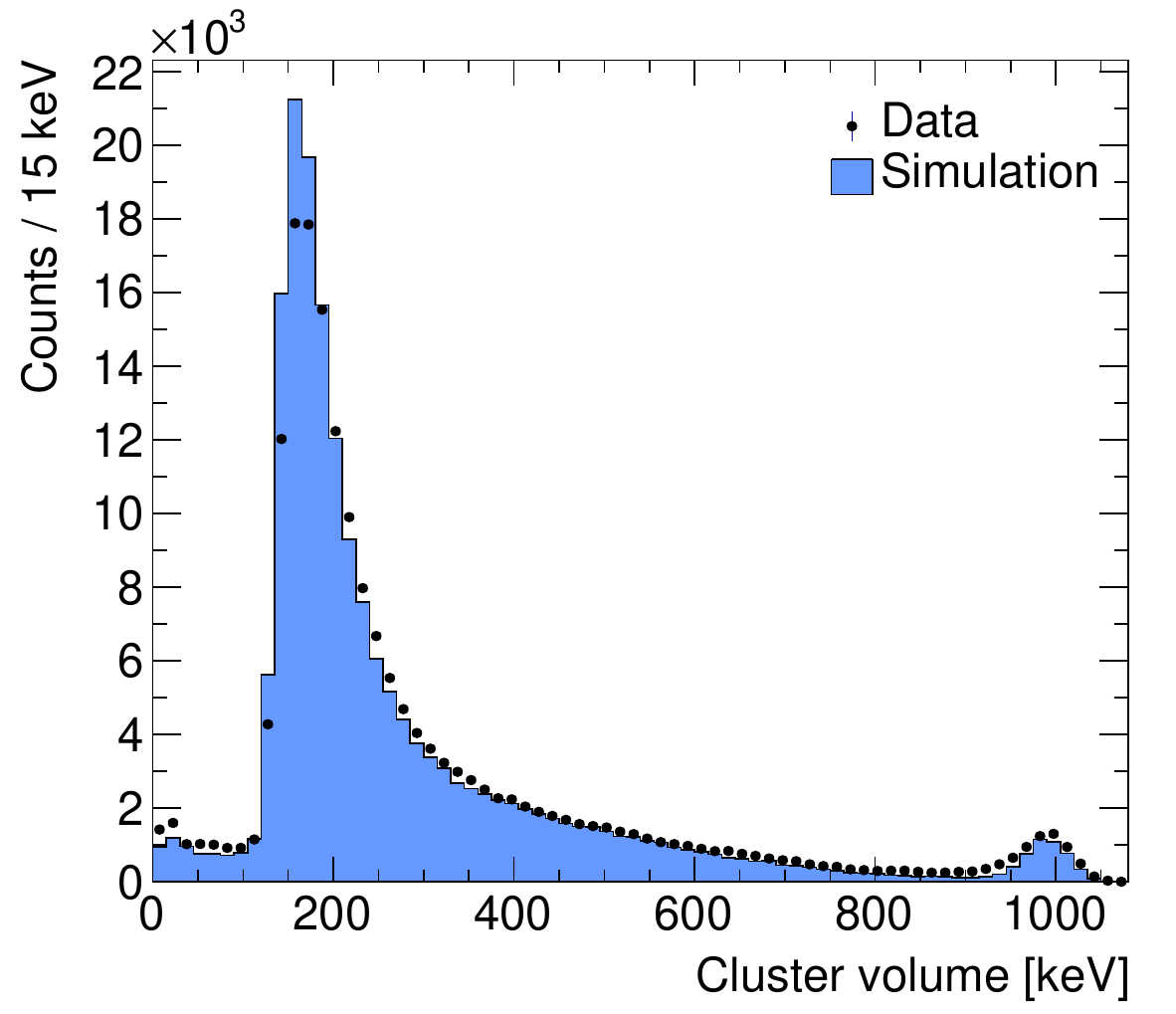}
  \caption{\label{fig:KralclstrSize} 
    Comparison of deposited-energy distributions for 1-MeV
    electrons in experiment and simulation. [Cluster volume
    is the sum of energies deposited in contiguous activated
    pixels (a cluster); an event can sometimes consist of
    several clusters (then they are counted separately),
    both in experiment and simulation, due to, e.g,
    insufficient energy deposition in some pixels, or also
    backscattering of particles from construction materials.]}
\end{figure}

Figure~\ref{fig:KralclstrSize} shows the experimental and
simulated distribution of energy deposited by 1 MeV
electrons in the sensor; the number of simulated and
experimental events were the same.  The experimental data
are well reproduced. Since the majority of electrons pass
through the sensor, the Landau-like curve is to be expected
and is indeed obtained, with the most probable energy loss
of 150 keV in accord with the sensor thickness. The smaller
peak at 1 MeV is a full-energy peak brought about by
electrons scattered enough times in the sensor to leave all
their kinetic energy there, cf. sec.~\ref{ssec:lin} and
Fig.~\ref{fig:lin} below.

\subsection{Spatial distribution}

\begin{figure}[htbp]
  \begin{subfigure}{0.44\textwidth}
    \includegraphics[width=\textwidth]{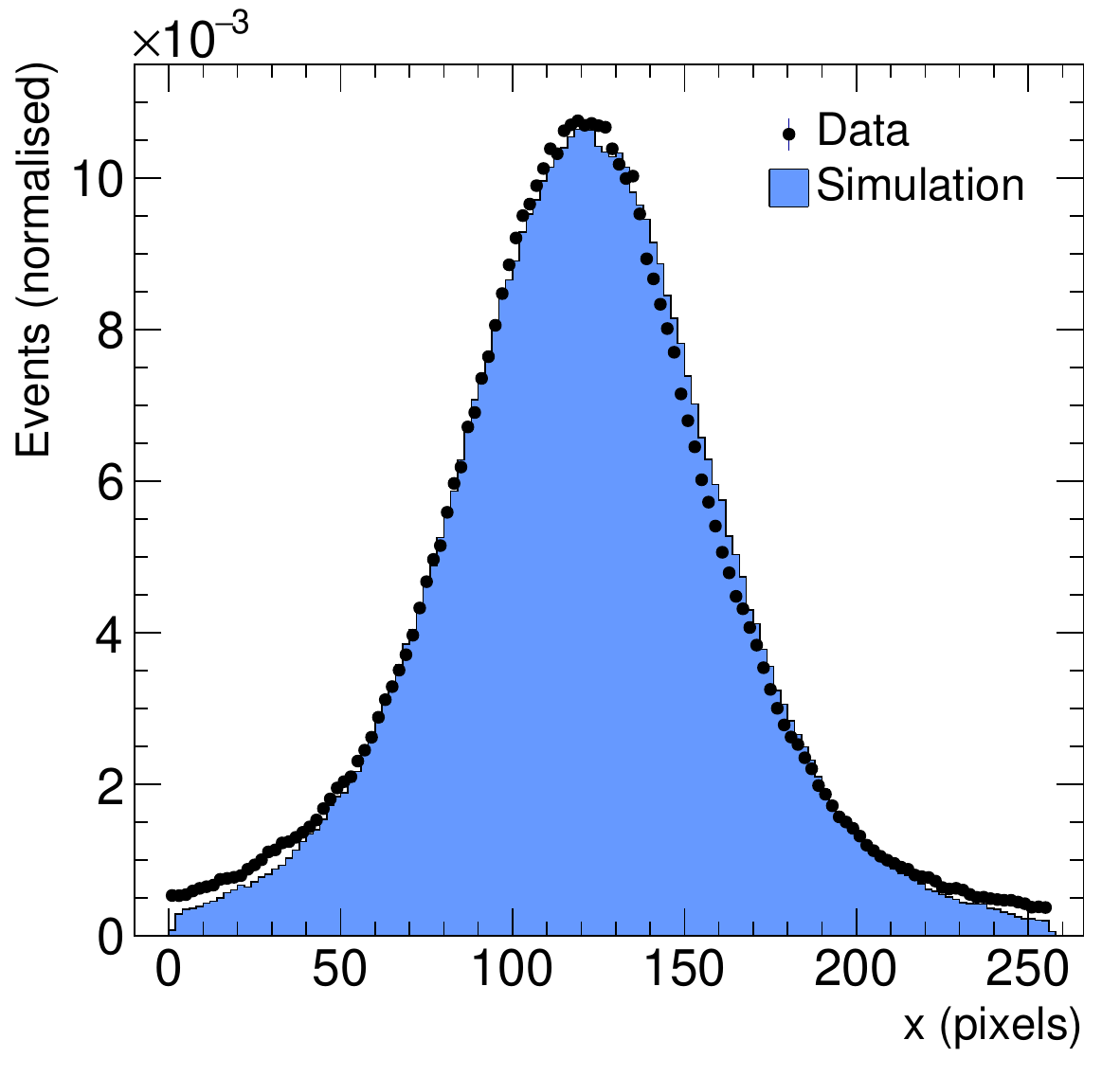}
    \caption{}
    \label{fig:KralXY_c}
  \end{subfigure}\qquad
  \begin{subfigure}{0.44\textwidth}
    \includegraphics[width=\textwidth]{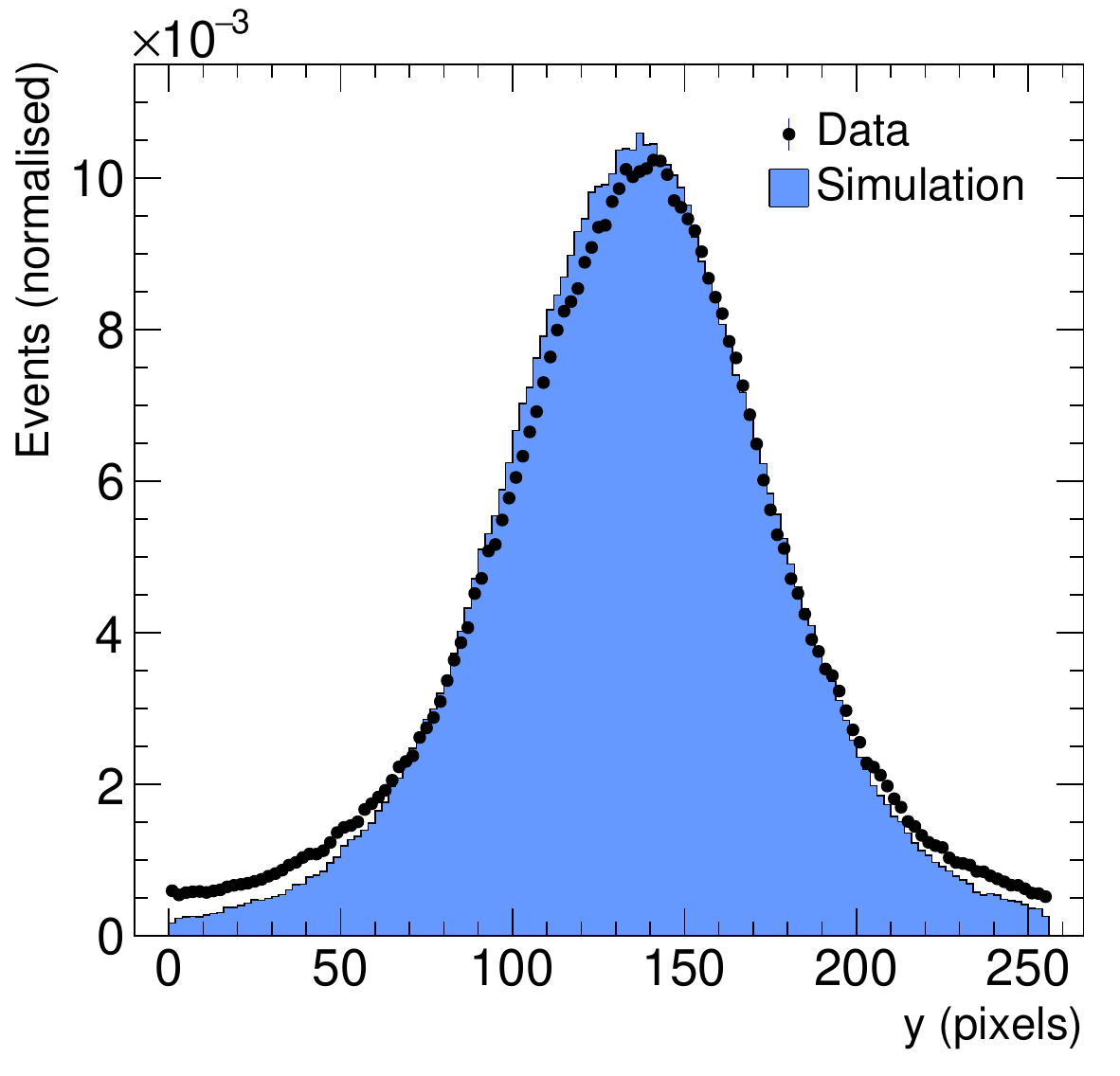}
    \caption{}
    \label{fig:KralXY_d}
  \end{subfigure}\qquad
  \caption{\label{fig:KralXY} The spatial distribution of
   activated TPX3 pixels in experiment and simulation along
   (a) the $x$-axis and (b) the $y$-axis.}
\end{figure}

The spatial distributions of activated pixels (hits) along
the $x$ and $y$ axes are depicted in
figure~\ref{fig:KralXY_c} and~\ref{fig:KralXY_d},
respectively.  The simulation accurately reproduces the
distribution along the $x$ axis (along which the electron
movement is not affected by the magnetic field), and
slightly less so along the $y$ axis, namely in the tails of
the distribution.  To rule out detector-related effects as
the cause of the discrepancy we compared the above
distributions, counting all activated pixels, with analogous
spatial distributions of the first hit of each detected
track (entry point to the detector; pixel with maximum ToA).
The resulting distributions were essentially identical to
those in Fig.~\ref{fig:KralXY}, hence the discrepancy is not
caused by what happens in the detector. It may, however, be
related to the differences between the actual experimental
setup and its simulated model, as discussed in
Section~\ref{sec:simul}.

\subsection{Track linearity}
\label{ssec:lin}

The track linearity is defined as the fraction of activated
pixels intersected by a straight line drawn between a
track's start and end points. I.e., it is one for straight
tracks and generally decreases when scattering (change in
direction) occurs along the track; linear tracks are more
probable for particles of higher energies.

\begin{figure}[htbp]
	\centering 
        \begin{subfigure}{0.45\textwidth}
          \includegraphics[height=4.5cm]{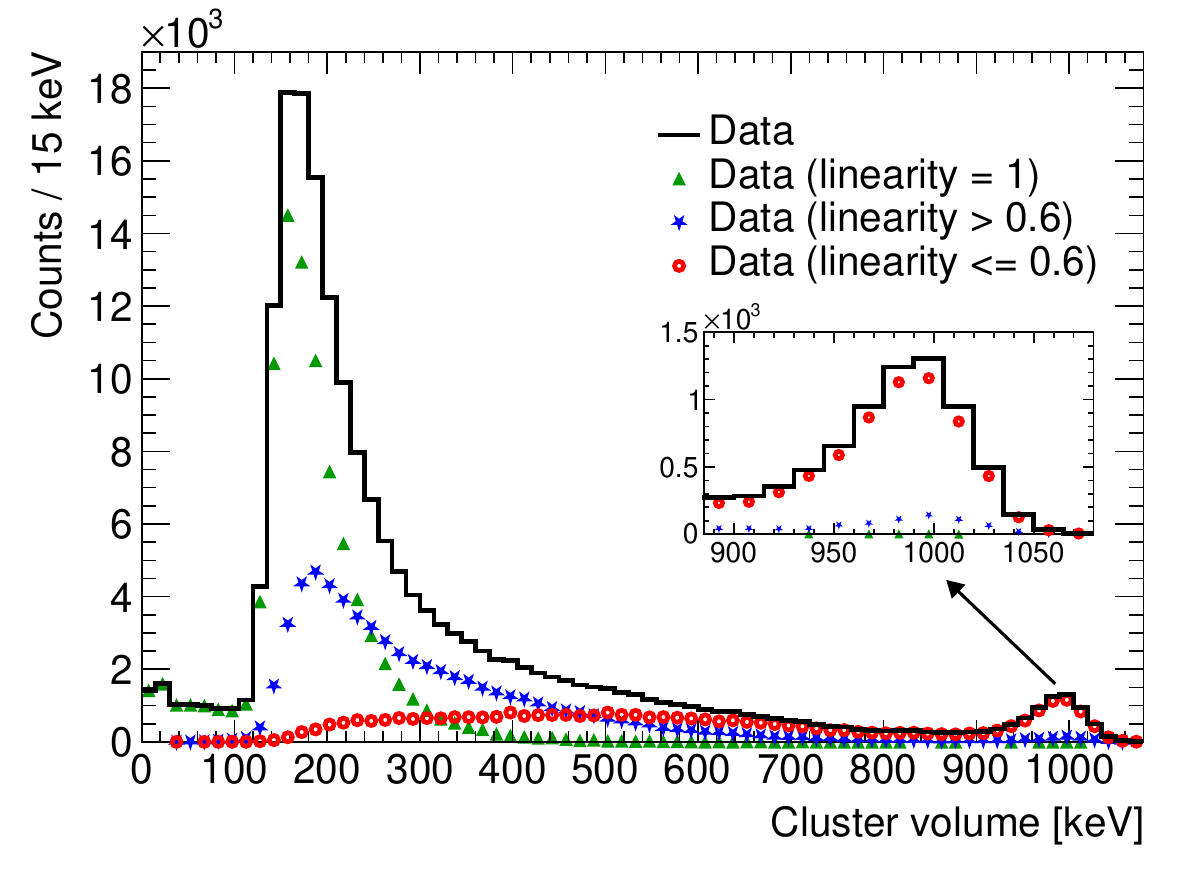}
          \caption{}
          \label{fig:lin_a}
        \end{subfigure}\qquad
        \begin{subfigure}{0.45\textwidth}
          \includegraphics[height=4.5cm]{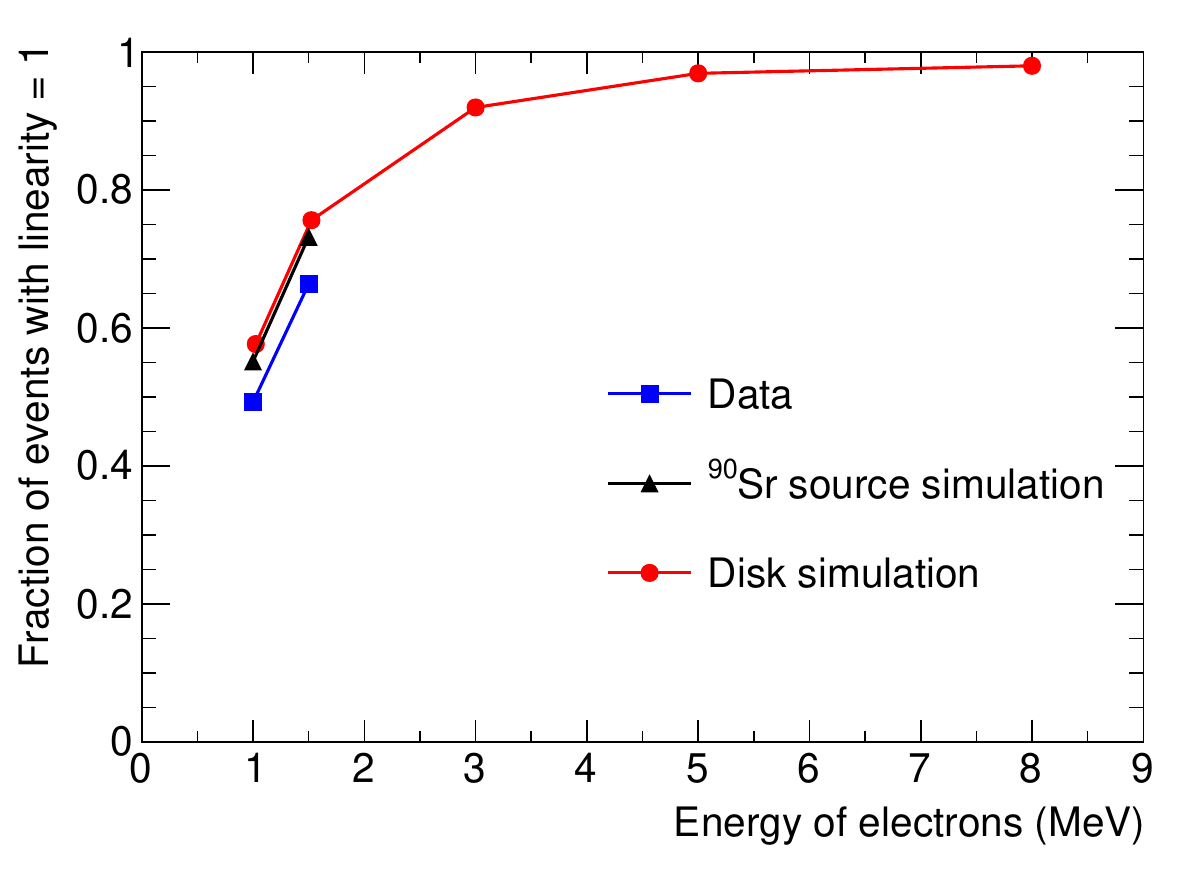}
          \caption{}
          \label{fig:lin_b}
        \end{subfigure}
        \caption{\label{fig:lin} (a) Experimental deposited-energy spectrum
        for 1-MeV electrons with tracks categorized into three groups based on
	their linearity.
        (b) The fraction of perfectly linear tracks (linearity equal
        to 1) for different electron energies.}
\end{figure}

Figure \ref{fig:lin_a} displays the (experimental)
correlation of track linearity with the energy deposited by
1-MeV electrons in the sensor. As mentioned in section
\ref{ssec:edist}, 1-MeV electrons have the range well
exceeding the sensor thickness.  Hence the energy deposited
by electrons that do not undergo substantial scattering --
their tracks have high linearity -- should follow the Landau
curve. On the other hand, electrons with tracks of low
linearity should contribute to the full-energy peak. The
figure classifies (somewhat arbitrarily) the tracks into
three groups according to their linearity, and confirms the
argument.

Besides electrons of 1\,MeV also 1.5-MeV ones were measured.
Blue points (squares) in figure \ref{fig:lin_b} then
substantiate the intuitive picture that linearity of tracks
grows with energy. For these two energies
whole-setup-simulations were carried out, resulting in the
black points (triangles) of the figure, in 10\% agreement
with the data. Due to limitations of the electron source, no
experimental data is available for higher energies, and only
simulation results, here for energies up to 8\,MeV, can be
shown. Furthermore, since simulation of the whole setup is
computationally rather intensive, what is displayed as the
figure's red points (circles) are results of our older
simplified simulations, in which the whole electron-source
setup was replaced with a a point source emitting electrons
of a given energy towards the detector, with a Gaussian-like
distribution of directions deflecting from the shortest line
from the source to the detector centre. To reproduce the
experimental findings for 1 and 1.5-MeV electrons, the point
source was placed 23 mm in front of the detector and the
sigma of the polar-angle distribution was set to 4.5$^\circ$.
(We note that although in the experiment there is 36 mm of
air between the particle source and the detector, leading to
non-negligible multiple scattering, we had vacuum there in
this simplified simulation, and the effect of the scattering
was, effectively, incorporated into the sigma and the
shorter distance to the detector.)

From Fig.~\ref{fig:lin_b} it follows that, in a 0.5-mm-thick
silicon sensor, above 95\% of electrons with energy above
5\,MeV have their track linearity equal to 1.

\section{Conclusion}
\label{sec:conclusion}

Semi-relativistic 0.5 and 1.5\,MeV electrons were measured
with a Timepix3 detector using a 0.5-mm Si sensor. The
measurement was compared with a full-setup simulation, which
modelled both the inner parts of the monochromator that
selected electrons of a given energy and the processes of
the detection proper inside the sensor. The agreement
between the simulation and experiment, concerning the
deposited-energy spectra, spatial profiles of the electron
beam, and tracks linearity is satisfactory in general. The
work brought some understanding of how linearity of tracks
in the used Si sensor evolves with energy. In particular,
one can expect that already at 3\,MeV over 90\% of electrons
(positrons) leave tracks of linearity (of our definition)
equal to one.     
 
The results indicate that the developed simulation framework
is trustworthy and can be used for our forthcoming studies
engaging higher-energy electrons and positrons related to
the ATOMKI-anomaly.


\acknowledgments

Funding: This work was supported by GA\v CR -- Czech Science
Foundation, grant GA21-21801S.

Computational resources were provided by the e-INFRA CZ
project (ID:90140), supported by the Ministry of Education,
Youth and Sports of the Czech Republic.

The authors would like to thank Petr Smolyanskyi (IEAP CTU)
for the support in the calibration of the Timepix3
detectors.



 \bibliographystyle{JHEP}
 \bibliography{refs}

\end{document}